# A SEMI-EMPIRICAL BAYESIAN CHART TO MONITOR WEIBULL PERCENTILES[*]

PASQUALE ERTO[¶], GIULIANA PALLOTTA[¶] AND CHRISTINA M. MASTRANGELO[†]

[¶]UNIVERSITY OF NAPLES "FEDERICO II"
P.LE V. TECCHIO 80, 80125 NAPLES
ITALY

[†]UNIVERSITY OF WASHINGTON
SEATTLE, WA, 98195
USA

This paper develops a Bayesian control chart for the percentiles of the Weibull distribution, when both its in-control and out-of-control parameters are unknown. The Bayesian approach enhances parameter estimates for small sample sizes that occur when monitoring rare events as in high-reliability applications or genetic mutations. The chart monitors the parameters of the Weibull distribution directly, instead of transforming the data as most Weibull-based charts do in order to comply with their normality assumption. The chart uses the whole accumulated knowledge resulting from the likelihood of the current sample combined with the information given by both the initial prior knowledge and all the past samples. The chart is adapting since its control limits change (e.g. narrow) during the Phase I. An example is presented and good Average Run Length properties are demonstrated. In addition, the paper gives insights into the nature of monitoring Weibull processes by highlighting the relationship between distribution and process parameters.

## 1. Introduction

Statistical process monitoring techniques have traditionally gained a wide acceptance among design and process engineers to verify the quality of their products. However, during the last decades, these techniques found many other applications such as health care monitoring, detection of genetic mutation, credit card fraud detection and insider trading in stock markets, as reported in Chatterjee and Qiu (2009) and the references therein. So their applications have now moved far beyond engineering into management, environmental science, molecular biology, genetics, epidemiology, clinical medicine, finance, law enforcement, athletics (Stoumbos et al. 2000) and even into monitoring the payments made to welfare recipients (Fairley et al. 1990).

When a key quantity such as the life length of whichever thing must be monitored, the solution is a control chart of a specific characteristics associated with the key quantity, whose variability is often modelled via skewed distribution as the Weibull one (Meeker and Hamada, 1995). The Weibull distribution has found a myriad of different applications such as to predict prognosis for gastric cancer patients (Miceli et al. 2004), to model the survival times for the Acacia mangium plantation (Kudus et al. 2006) or the time between ozone exceed-

---

ances of a given threshold in some regions (Achcar et al. 2012). In general the Weibull distribution has been used in many areas, such as medicine, engineering, biology, public health, epidemiology, economics, demography, criminology, sociology, and ecology.

However, the Weibull distribution is more widely used to describe a random variable such as the time to failure of a mechanical device or the strength of brittle materials. As an example, in the manufacturing process of materials, breaking strength is monitored as it is directly associated with the material performance. To perform quality control, samples are randomly selected from the production process and sent to a destructive test, where the time to failure or the breaking stress is measured. The Weibull distribution is considered to adequately fit such reliability data. Generally, that results from the flexibility of its hazard rate; specifically, that can be also justified in the frame of the Extreme Value Theory, where the Weibull is one of the three asymptotes: the failure occurs at the weakest flaw [see, e.g., Padgett and Spurrier (1990), Pascual (2010)].

Rapid innovation and intense competition require failure data checks repeatedly; however, collecting failure data is usually expensive and time consuming. Consequently, a dramatic reduction of the amount of data needed for reliability monitoring is largely desirable.

Most of the related literature in reliability monitoring focuses on the time between failures [Xie et al. (2002), Sürücü and Sazak (2009)], which uses control charts based on the Exponential and the Gamma distributions. Batson et al. (2006) provide guidelines to use Shewhart control charts for monitoring field failure data, making use of normalizing transformations and highlighting the issues arising from the use of small samples.

Classical Shewhart control charts for mean (and variance) under normality assumption are not effective when the distributions of reliability parameters are skewed, if the small size of the available samples prevents the use of the "normalizing effect" for the sample mean [see Shore (2004) and the references therein]. The case of the (skewed) Log-Normal distribution is an exception since the logarithm transformation of the observed data normalize them. Padgett and Spurrier (1990), however, do propose Shewhart-type control charts for transformed percentiles of the Weibull and log-normal distributions for reliability monitoring. Kanji and Arif (2001) also consider the development of a control chart using a percentile approach for Weibull data. Zhang and Chen (2004) investigate alternative charts to detect changes (both decreases and increases) in the process mean for censored Weibull lifetimes. Recently, Pascual (2010), Pascual and Nguyen (2011) and Pascual and Li (2012) have shown that the sample mean is not appropriate to monitor the Weibull data because the central limit theorem does not hold. In Pascual (2010) an Exponentially Weighted Moving Average (EWMA) control chart is proposed to check the stability of the Weibull shape parameter by transforming the Weibull into the Smallest Extreme Gumbel distribution.

When we have to face estimation procedures based on extremely small samples, the use of Bayesian methods appears appropriate since alternative classical estimators, like the Maximum Likelihood ones, give estimates that are very often worse than and/or in contrast with elementary scientific or technological knowledge, as shown, for instance, in Canavos and Tsokos (1972 and 1973), Smith and Naylor (1987) and Erto (2005). On the other hand, in technological fields such as engineering, the Bayes method can usually rely on available prior information. In fact, in engineering always some knowledge exists about the mechanism of failure under consideration, which can be converted into quantitative form about the Weibull parameters.

A survey on Bayesian statistics in process monitoring, control and optimization is given by Colosimo and Del Castillo (2007) and Colosimo (2008). Some Bayesian-type charts are presented in Tsiamyrtzis and Hawkins (2005, 2007, 2008 and 2010), and in Zamba et al. (2008), to detect changes in the mean of a Normal distribution.

The present paper starts from the results in Erto and Pallotta (2007), where *static* Shewhart-type control chart is used to monitor the percentiles of the Weibull distribution. That control chart already employs Bayesian estimators for Weibull parameters and shows a detection power higher among others for Weibull processes as demonstrated in [Hsu et al. (2011)]. However, it differs from the one proposed here as it is memory-less, since the Bayes' theorem combines the given initial prior information exclusively with the *current* sample data, at each sampling stage. Moreover the chart uses *fixed* control limits, i.e. there is no updating of control limits from period to period. Instead, the present paper shows two new features: a) the use of a Bayesian approach that continuously combines prior information with *all* accumulated past and present data in order to perform a "learning by sampling"; b) control limits which are *continuously refined* during the Phase I.

We call Phase I analysis the retrospective analysis usually applied to a sufficiently great number (say 50) of historical in-control process data (often called *training* data), in order to identify the in-control process parameters as accurately as possible, to allow quality engineers to design the control charts for a future process monitoring.

Mathematically, feature a) leads to replicated priors. A first practical consequence is that the points on the chart are *cumulative* estimates, instead of *single-sample* estimates. So, the proposed chart is memory-type. The feature b) provides the chart with control limits that can be calculated since the very first sample, making the proposed chart a potential candidate in short-runs and low-volume production settings, as well as it provides the chart with a quick start.

It is interesting to note that the proposed chart has the potential to be applied as an acceptance control chart straightforward. This special kind of chart samples from a lot and not from the process, and usually are employed to control a feature of the product on the basis of some specification limits. Thus, differently from traditional control charts, the feature of the product is allowed to drift while it is still inside the specification limits. This is due to some particular technological environments, such as chemical processes or tool wear, where some features of the product may follow a controlled drift (see, e.g., Wesolowky ,1990). The main requirement is that the process dispersion must be sufficiently small, compared to the specification limits (i.e., the process capability must be high). In these cases, a common control chart cannot be applied since, when the process changes, the *specific* feature of the product could be still inside the specifications.

The remainder of the paper is organized as follows. The next section introduces the Weibull distribution, and the practical importance of the Weibull parameters is discussed. The following section describes the details of the primary steps to design and develop the chart to monitor Weibull percentiles. An example is reported in the next section, using data published in Padgett and Spurrier (1990). Finally the results of a sensitivity/performance investigation are given.

## 2. Using the Weibull Percentiles for Reliability Monitoring

The Weibull distribution is widely-used in reliability testing, survival analysis and quality control, since it possesses many useful properties: it has only two parameters; its form is flexible to model different shapes; it has a simple likelihood function. The breaking strengths of brittle materials, the lifetimes of electronic devises and the survival times of patients are typical examples of random variables having a Weibull distribution. Given a random variable $x$, the corresponding Weibull cumulative density function (Cdf) is:

$$F(x;\delta,\beta) = 1 - \exp\left[-(x/\delta)^\beta\right] \qquad x \geq 0; \quad \delta,\beta > 0 \qquad (1)$$

where $\delta$ is the scale parameter and $\beta$ is the shape parameter.

Let $R$ denote a specified reliability level. Then, the corresponding Weibull percentile $x_R$ can be expressed as $x_R = \delta \left[\ln(1/R)\right]^{1/\beta}$ by setting (1) equal to $1-R$ and solving for $x_R$.

In materials science, the monitoring of a specific Weibull percentile, $x_R$, becomes critical when the quality characteristic is the strength or the time to failure of brittle and quasi-brittle materials. In these contexts, monitoring the process mean and variance by means of classical control charts is less effective than monitoring percentiles since a small variation in mean and/or variance can produce a significant shift in $x_R$ (see Padgett and Spurrier, 1990). Within this framework, $x_R$ is considered a minimum threshold for reliability design and is more critical than the scale parameter $\delta$ or the mean $\mu$ (see, e.g., Bao et al., 2007). On the other side, the Weibull shape parameter, $\beta$, is a characteristic of the phenomenon under consideration. In fact it is related to the dispersion of flaws in the material (e.g. see Padgett et al. 1995). Thus $\beta$ can be considered constant, even if unknown as, for example, in Nelson (1979). Specifically, when flaws are evenly distributed throughout the volume, the material is more consistent, and the Weibull shape parameter is larger. As such, a material with a large $\beta$ shows lower variability, while a material with a small $\beta$ shows a higher variability in strength (due to clusters of flaws). For most brittle materials, the $\beta$ value is likely included in the range between 3 and 10.

Also in survival analysis, the Weibull shape parameter is usually constant and the prior knowledge for $\beta$ may use information about the failure mechanism: infant mortality failures imply $\beta < 1$; (e.g., when using new technology); random failures imply $\beta = 1$ (e.g., due to external shocks); and wear out failures imply $\beta > 1$ (e.g., fatigue cracks) [see, e.g., Meeker and Hamada (1995)].

Moreover, an engineer presumably knows more than the simple order of magnitude of the performance (e.g., life length or breaking strength) that the item he designed has; therefore, he has a quite precise knowledge of a percentile [Erto (1982)]. On the other side, an engineer is not used to express his prior knowledge in terms of the scale parameter $\delta$ but in terms of the less alien and more practical concept of the Weibull percentile $x_R$.

Generally speaking, in reliability engineering there is always prior information which can be reasonably quantified in terms of: 1) a range $(\beta_1, \beta_2)$ of the shape parameter, and 2) an anticipated value $\bar{x}_R$ for a percentile of the sampling distribution [Erto (1982)].

For all these reasons, the re-parameterized form of Cdf (1) in terms of the percentile $x_R$ and the shape parameter $\beta$:

$$F(x; x_R, \beta) = 1 - \exp\left[-\ln(1/R)(x/x_R)^\beta\right]; \quad x \geq 0; \quad x_R, \beta > 0; \quad (2)$$

is adopted, since it is a more suitable basis for the control chart developed in the next section.

## 3. The Design and Development of the Control Chart

### 3.1 *Initialization*

The first step in the design of the proposed control chart is to specify a reliability level $R$ of interest. In the literature, $R$ is usually in the 90 - 99% range, since a high reliability level is generally required in most industries [Meeker and Hamada (1995)]. The value of $R$ is usually specified as the result of a design process.

One advantage of the proposed Bayesian approach is that it suffices to anticipate a numerical interval $(\beta_1, \beta_2)$ for the shape parameter $\beta$ and a likely value for the percentile $x_R$, based on previous experiments and expert opinion.

For $\beta$ the Uniform prior pdf is assumed in the interval $(\beta_1, \beta_2)$ since this model fits well the degree of belief and it appears to be as non-restrictive as feasible. The interval $(\beta_1, \beta_2)$ must be chosen wide enough in order to plausibly contain the unknown (true) value of the Weibull shape parameter. In fact, outside the $(\beta_1, \beta_2)$ interval we set zero prior probability and thus the posterior distribution of the unknown parameter $\beta$ will never be able to get values outside this pre-specified interval, both in-control and auto-of-control states.

The prior probability density function of $x_R$ is assumed to be the Inverse Weibull since the distribution fits well with the prior knowledge about $x_R$ and is tractable by practitioners [Erto (1982)]:

$$\text{pdf}\{x_R\} = a b (a x_R)^{-(b+1)} \exp\left[-(a x_R)^{-b}\right]; \quad x_R \geq 0; \quad a, b > 0 \quad (3)$$

where $a$ and $b$ are the scale and shape parameters respectively.

We note that the greater the shape parameter $\beta$ is, the more peaked the Weibull pdf is, the smaller the uncertainty in $x_R$ is, and then greater $b$ must be. Therefore, the simplest choice is assuming $b = \beta$, as suggested in Erto (1982). This assumption leaves $b$ unspecified, and the practice shows that it works and it is preferable rather than to choose a fixed value, say $b = 3$, which often includes other information that is not actual prior knowledge about the prior.

The expected value of (3) is:

$$E\{x_R\} = \frac{\Gamma(1 - 1/b)}{a}. \quad (4)$$

The scale parameter $a$ can be then evaluated as:

$$a = \frac{\Gamma\left(1 - 1/\bar{b}\right)}{\bar{x}_R} \tag{5}$$

where $\bar{x}_R$ is the anticipated value for $E\{x_R\}$, obtained from the prior information about $x_R$, and $\bar{b}$ is the value of $b$ that we can anticipate on the basis of the prior interval $(\beta_1, \beta_2)$ for $\beta$ (coherently with the above assumption $b = \beta$):

$$\bar{b} = (\beta_1 + \beta_2)/2. \tag{6}$$

The only restriction is $\beta_1 + \beta_2 > 2$ since $\beta_1 + \beta_2$ is used to set up the argument of the Gamma function in (4). This means that we cannot set both values less than the unit even if, for instance, we are concerned with infant mortality failures that imply $\beta > 1$; so, if we decide to anticipate $\beta_1 = 0.1$ we are obliged to set $\beta_2 > 1.9$. However, anticipating a so wide interval does not affect significantly the properties of the Bayes estimators since, as said before, the most important thing is that the interval must contain the unknown (true) value of $\beta$.

The joint prior probability density function (pdf) of $x_R$ and $\beta$ is:

$$\text{pdf}\{x_R, \beta\} = (\beta_2 - \beta_1)^{-1} a \beta (a x_R)^{-(\beta+1)} \exp\left[-(a x_R)^{-\beta}\right] \tag{7}$$

which combines the Uniform prior pdf for $\beta$ and the Inverse Weibull prior pdf for $x_R$ in (3). Then this prior information about $x_R$ and $\beta$ can be incorporated into a Bayesian monitoring scheme.

### 3.2  *Development of the Posterior Distribution*

Consider the first random sample vector $\underline{x}_1$ of $n$ data. The corresponding likelihood function, under the Weibull assumption, is:

$$L\left(\underline{x}_1 | x_R, \beta\right) \propto \beta^n x_R^{-\beta n} \prod_{i=1}^{n} x_i^{\beta-1} \exp\left[-x_R^{-\beta} \ln\left(\frac{1}{R}\right) \sum_{i=1}^{n} x_i^{\beta}\right] \tag{8}$$

The Bayes' theorem combines the prior distribution (7) with the likelihood function (8) to obtain the following joint posterior pdf:

$$\text{pdf}\{x_R, \beta | \underline{x}_1\} = \frac{\text{pdf}\{x_R, \beta\} L\left(\underline{x} | x_R, \beta\right)}{\int_{\beta_1}^{\beta_2} \int_0^{\infty} \text{pdf}\{x_R, \beta\} L\left(\underline{x}_1 | x_R, \beta\right) dx_R d\beta} =$$

$$= \frac{\beta^{n+1} a^{-\beta} x_R^{-\beta(n+1)-1} \prod_{i=1}^{n} x_i^{\beta-1} \exp\left[-x_R^{-\beta}\left(a^{-\beta} + \ln\left(\frac{1}{R}\right)\sum_{i=1}^{n} x_i^{\beta}\right)\right]}{\Gamma(n+1) \int_{\beta_1}^{\beta_2} \beta^n a^{-\beta} \prod_{i=1}^{n} x_i^{\beta-1} \left(a^{-\beta} + \ln\left(\frac{1}{R}\right)\sum_{i=1}^{n} x_i^{\beta}\right)^{-(n+1)} d\beta}. \tag{9}$$

Consider the second random sample vector $\underline{x}_2$ of $n$ data. Combine the corresponding likelihood function:

$$L\left(\underline{x}_2 | x_R, \beta\right) \propto \beta^n x_R^{-\beta \cdot n} \prod_{i=n+1}^{2 \cdot n} x_i^{\beta-1} \exp\left[-x_R^{-\beta} \ln\left(\frac{1}{R}\right) \sum_{i=n+1}^{2 \cdot n} x_i^{\beta}\right] \quad (10)$$

with the posterior distribution (9) (used as the prior for the current sample) and obtain the second joint posterior pdf:

$$\text{pdf}\left\{x_R, \beta | \underline{x}_2\right\} = \frac{\beta^{2 \cdot n+1} a^{-\beta} x_R^{-\beta(2 \cdot n+1)-1} \prod_{i=1}^{2 \cdot n} x_i^{\beta-1} \exp\left[-x_R^{-\beta}\left(a^{-\beta} + \ln\left(\frac{1}{R}\right) \sum_{i=1}^{2 \cdot n} x_i^{\beta}\right)\right]}{\Gamma(2 \cdot n+1) \int_{\beta_1}^{\beta_2} \beta^{2 \cdot n} a^{-\beta} \prod_{i=1}^{2 \cdot n} x_i^{\beta-1} \left(a^{-\beta} + \ln\left(\frac{1}{R}\right) \sum_{i=1}^{2 \cdot n} x_i^{\beta}\right)^{-(2 \cdot n+1)} d\beta} \quad (11)$$

Note that at each sample, the joint posterior pdf includes the whole accumulated dataset, i.e. all the samples of size $n$ taken until then. Thus, recursively, at the $k^{\text{th}}$ sample $\underline{x}_k$, the posterior pdf of $x_R$ and $\beta$ is:

$$\text{pdf}\left\{x_R, \beta | \underline{x}_k\right\} = \frac{\beta^{k \cdot n+1} a^{-\beta} x_R^{-\beta(k \cdot n+1)-1} \prod_{i=1}^{k \cdot n} x_i^{\beta-1} \exp\left[-x_R^{-\beta}\left(a^{-\beta} + \ln\left(\frac{1}{R}\right) \sum_{i=1}^{k \cdot n} x_i^{\beta}\right)\right]}{\Gamma(k \cdot n+1) \int_{\beta_1}^{\beta_2} \beta^{k \cdot n} a^{-\beta} \prod_{i=1}^{k \cdot n} x_i^{\beta-1} \left(a^{-\beta} + \ln\left(\frac{1}{R}\right) \sum_{i=1}^{k \cdot n} x_i^{\beta}\right)^{-(k \cdot n+1)} d\beta} \quad (12)$$

where the value of the prior parameter $a$ is obtained by replacing $\bar{x}_R$ in (5) with $\hat{x}_{R,k-1}$, the point estimate (16) of the $x_R$ percentile from the previous $(k-1)^{\text{th}}$ sample. Beside, if the estimate (14) of the shape parameter $\beta$ from the previous $(k-1)^{\text{th}}$ sample is greater than the unit, it is enough to adopt $\beta_1 = \hat{\beta}_{k-1}/2$ and $\beta_2 = \hat{\beta}_{k-1} \times 1.5$ in order to obtain a reasonable large symmetrical interval that complies with the restriction $\beta_1 + \beta_2 > 2$.

The equation (12) represents both the posterior distribution after the $k^{\text{th}}$ sample and the prior for the $(k+1)^{\text{th}}$ one. These replicated distributions stand for all sampling stages $(k \geq 1)$, but differ from the initial prior (7) $(k = 0)$.

In general, the posterior density function (12) provides the opportunity to continually update the parameters as soon as a new sample becomes available. It describes the residual uncertainty, i.e., the uncertainty that remains about the Weibull parameters after the $k^{\text{th}}$ sampling.

Note that the equation (12) is different from that proposed in Erto and Pallotta (2007) and identically reported in Hsu et al. (2011).

### 3.3 *Obtain Posterior Parameter Estimates*

After the $k^{\text{th}}$ sample, from equation (12) we can obtain the marginal posterior pdf for $\beta$, $\text{pdf}\{\beta | \underline{x}_k\}$, by integrating over $x_R$:

$$\text{pdf}\{\beta | \underline{x}_k\} = \frac{\beta^{k \cdot n} a^{-\beta} \prod_{i=1}^{k \cdot n} x_i^{\beta-1} \left( a^{-\beta} + \ln\left(\frac{1}{R}\right) \sum_{i=1}^{k \cdot n} x_i^{\beta} \right)^{-(k \cdot n + 1)}}{\int_{\beta_1}^{\beta_2} \beta^{k \cdot n} a^{-\beta} \prod_{i=1}^{k \cdot n} x_i^{\beta-1} \left( a^{-\beta} + \ln\left(\frac{1}{R}\right) \sum_{i=1}^{k \cdot n} x_i^{\beta} \right)^{-(k \cdot n + 1)} d\beta}. \quad (13)$$

Thus, the estimate of the shape parameter $\beta$ is the posterior expectation based on (13):

$$\hat{\beta}_k = \text{E}\{\beta | \underline{x}_k\} = \frac{\int_{\beta_1}^{\beta_2} \beta^{k \cdot n + 1} a^{-\beta} \prod_{i=1}^{k \cdot n} x_i^{\beta-1} \left( a^{-\beta} + \ln\left(\frac{1}{R}\right) \sum_{i=1}^{k \cdot n} x_i^{\beta} \right)^{-(k \cdot n + 1)} d\beta}{\int_{\beta_1}^{\beta_2} \beta^{k \cdot n} a^{-\beta} \prod_{i=1}^{k \cdot n} x_i^{\beta-1} \left( a^{-\beta} + \ln\left(\frac{1}{R}\right) \sum_{i=1}^{k \cdot n} x_i^{\beta} \right)^{-(k \cdot n + 1)} d\beta} \quad (14)$$

The above assumption $b = \beta$ reinforces the natural mathematical dependency of the percentile $x_R$ on the parameter $\beta$. So, being $x_R$ highly dependent on $\beta$, a conditional posterior pdf for $x_R$ must be considered to improve the responsiveness of the chart and then its out-of-control detection property:

$$\text{pdf}\{x_R | \underline{x}_k, \beta\} = \frac{\text{pdf}\{x_R, \beta | \underline{x}_k\}}{\text{pdf}\{\beta | \underline{x}_k\}} =$$

$$= \frac{\beta x_R^{-\beta(k \cdot n + 1) - 1}}{\Gamma(k \cdot n + 1)} \left( a^{-\beta} + \ln\left(\frac{1}{R}\right) \sum_{i=1}^{k \cdot n} x_i^{\beta} \right)^{k \cdot n + 1} \exp\left[ -x_R^{-\beta} \left( a^{-\beta} + \ln\left(\frac{1}{R}\right) \sum_{i=1}^{k \cdot n} x_i^{\beta} \right) \right] \quad (15)$$

The point estimate of the $x_R$ percentile is the posterior expectation of (15):

$$\hat{x}_{R,k} = \text{E}\{x_R | \underline{x}_k, \overline{\beta}_k\} = \frac{\Gamma(k \cdot n + 1 - \overline{\beta}_k^{-1})}{\Gamma(k \cdot n + 1)} \left( a^{-\overline{\beta}_k} + \ln\left(\frac{1}{R}\right) \sum_{i=1}^{k \cdot n} x_i^{\overline{\beta}_k} \right)^{\frac{1}{\overline{\beta}_k}} \quad (16)$$

where:

$$\overline{\beta}_k = \frac{1}{k} \sum_{i=1}^{k} \hat{\beta}_i \quad (17)$$

is the average of all the posterior estimates from (14) accumulated up to and including the $k^{\text{th}}$ one. This estimate complies with the stability feature of $\beta$, discussed above, and allows inserting a robust value for $\beta$ in (16). Moreover, it is conservative (i.e. robust against sporad-ic/isolated $\beta$ variation), giving more weight to past data, since the calculation for each $\hat{\beta}_k$ uses the information from *all* previous samples.

Note that the key estimator (16) conceptually differs from that proposed in Erto and Pallotta (2007) and reported in Hsu et al. (2011) which, moreover, is not expressible in closed form.

At this point the estimate of the Weibull parameter $\hat{x}_{R,k}$ is obtained and is plotted on the control chart. After the Phase I, these point estimates will be compared with the control limits obtained as in the next step. Note that the monitoring of Weibull percentiles is run with no transformation needed.

Starting from the same joint posterior distribution in (12), the monitoring of the shape parameter $\beta$ can be performed too. This preliminary check, for stability of the shape parameter, can be necessary whenever technological evidence exists about a suspect shift in the Weibull shape from its initial stable value. In fact, based on the estimate in (14), we can set up an additional control scheme to monitor the stability of the shape parameter $\beta$. This control chart can be built by the side of the control scheme for the percentile $x_R$.

In our case, the control of parameter $\beta$ is not strictly needed due to the mentioned stability feature of $\beta$. However, for illustrative purposes, we present in Figure 1, b) the control chart of $\beta$ for the dataset used in Padgett and Spurrier (1990), where a decrease in $\beta$ is simulated.

### 3.4 *Obtain Probability-Based Control Limits*

Given a false alarm risk $\alpha$, we can obtain the lower control limit (LCL) $x_{R,\alpha/2}$ and the upper control limit (UCL) $x_{R,1-\alpha/2}$ for $x_R$ as the $\alpha/2$-percentiles of the posterior conditional distribution of $x_R$ in (15). To this end, we note that the conditional random variable $x_R | \underline{x}_k, \bar{\beta}_k$ can be transformed into a standard Gamma random variable. More specifically using the transformation:

$$z = x_R^{-\bar{\beta}_k} \left( a^{-\bar{\beta}_k} + \ln\left(\frac{1}{R}\right) \sum_{i=1}^{k \cdot n} x_i^{\bar{\beta}_k} \right) \quad (18)$$

and the $\text{pdf}\{x_R | \underline{x}_k, \bar{\beta}_k\}$, obtained from (15) for $\beta = \bar{\beta}_k$, it results into a pdf of $z$:

$$\text{pdf}\{z\} = \frac{z^{\gamma-1}}{\Gamma(\gamma)} \exp(-z) \quad (19)$$

where $\gamma$ is equal to $k \cdot n + 1$ and changes at each sample, based on the amount of *all* the accumulated data up to and including the current sample.

Using the inverse of the transformation (18):

$$x_R = z^{-1/\bar{\beta}_k} \left( a^{-\bar{\beta}_k} + \ln\left(\frac{1}{R}\right) \sum_{i=1}^{k \cdot n} x_i^{\bar{\beta}_k} \right)^{1/\bar{\beta}_k} \quad (20)$$

we can estimate the control limits $\mathrm{LCL} = x_{R,\alpha/2}$ and $\mathrm{LCL} = x_{R,1-\alpha/2}$ as simple transformations of the percentiles, $z_{1-\alpha/2}$ and $z_{\alpha/2}$ respectively, of the standard Gamma (19). These percentiles vary from sample to sample and their values obtained at the end of the Phase I become the control limits for future use of the chart. Once the Phase I is finished, a point that plots outside the control limits signals an out-of-control state, and the analyst goes back to the last in control point to restart monitoring samples as soon as the process is restored.

It is important to note that the greater the number of the training data of the Phase I is the lower the *first* priors effects are. In fact, by accumulating training data the weight of the sampling information tends to overcome the *initial* prior information about $x_R$ and $\beta$. Consequently, as the number of the training data increases the control limits approach a stable in-control values that express only the sampling variation. Simultaneously, the greater the number of the training data is the stronger the last joint posterior (12) of the Phase I is since it includes the whole accumulated dataset; because this posterior is used as prior for the following sampling, if it is excessively *strong*, in Bayes sense, the responsiveness of the chart toward eventual out-of-control data decreases consequently. In conclusion, even if exploiting as many training data as possible allows to set up the narrowest possible control interval, this advantage decreases rapidly; vice-versa, including too much training data (in the last joint posterior (12) of the Phase I) compromises the detection properties of the chart at the beginning of its use; so, we recommend to exploit about the usual fifty training data that represents an acceptable compromise in the case of this chart too.

While the centreline (CL) does not have a specific role on the chart, it may be added to the control scheme, similarly to the idea developed in Mastrangelo and Brown (2000).

For the same risk $\alpha$, we can also obtain the lower control limit (LCL) $\beta_{\alpha/2}$ and the upper control limit (UCL) $\beta_{1-\alpha/2}$ for $\beta$ as the $\alpha/2$-percentiles of the posterior distribution of $\beta$ in (13). This time we don't need any transformation, since both the equations:

$$\int_0^{\beta_{\alpha/2}} \mathrm{pdf}\{\beta|\underline{x}_k\} d\beta = \alpha/2; \qquad \int_0^{1-\beta_{\alpha/2}} \mathrm{pdf}\{\beta|\underline{x}_k\} d\beta = 1-\alpha/2 \qquad (21)$$

can be easily solved by using a standard numerical method (we tested also the Wolfram Mathematica 8 "FindRoot" routine successfully).

### 3.5　*Example: Control of Breaking Strength*

This section shows how to set up and use the proposed control chart on the Weibull data given in Padgett and Spurrier (1990). The purpose of the example is to monitor breaking strength of carbon fibres at a 99% reliability level. This means that the parameter of interest is the 0.01 percentile of the Weibull distribution and will be denoted as $x_{0.99}$. This carbon fibre is used to manufacture composite materials which requires a breaking strength greater than 1.22 GPa (Giga-Pascals) with 99% probability $(x_{0.99} = 1.22)$. A sample of $n = 5$ fibbers, each 50 mm long, is selected from the manufacturing process periodically, and the breaking stress of each fibre is measured. The first ten samples (of size $n = 5$) in Table 1 are assumed to be in-control, on the basis of previous technological considerations.

Figure 1a) shows the Bayesian control chart obtained using the first ten samples reported in Table 1 as the Phase I. On the basis of the information provided in Padgett and Spurrier, (1990), the first prior interval $4.8 \times (1 \mp 0.5)$ (i.e.: 2.4, 7.2) for $\beta$ and the anticipated (mean) value 1.22 for $x_R$ are adopted. Given a Shewhart-type false alarm risk $\alpha = 0.27$ %, the UCL and the LCL are calculated by (20). The first ten points represent the percentile $\hat{x}_{0.99}$, recursively estimated by (16).

The next ten samples given in Table 1 represent an out-of-control (OOC) state, simulated by shifting the first percentile to 0.26 GPa from the original value of 1.22 Gpa. This percentile shift expresses process deterioration caused by a decrease in mean and an increase in variance.

Table 1. Breaking Stresses (GPa) of Carbon Fibres

| Sample | Stress | | | | |
|---|---|---|---|---|---|
| 1  | 3.70 | 2.74 | 2.73 | 2.50 | 3.60 |
| 2  | 3.11 | 3.27 | 2.87 | 1.47 | 3.11 |
| 3  | 4.42 | 2.41 | 3.19 | 3.22 | 1.69 |
| 4  | 3.28 | 3.09 | 1.87 | 3.15 | 4.90 |
| 5  | 3.75 | 2.43 | 2.95 | 2.97 | 3.39 |
| 6  | 2.96 | 2.53 | 2.67 | 2.93 | 3.22 |
| 7  | 3.39 | 2.81 | 4.20 | 3.33 | 2.55 |
| 8  | 3.31 | 3.31 | 2.85 | 2.56 | 3.56 |
| 9  | 3.15 | 2.35 | 2.55 | 2.59 | 2.38 |
| 10 | 2.81 | 2.77 | 2.17 | 2.83 | 1.92 |
| 11 | 1.41 | 3.68 | 2.97 | 1.36 | 0.98 |
| 12 | 2.76 | 4.91 | 3.68 | 1.84 | 1.59 |
| 13 | 3.19 | 1.57 | 0.81 | 5.56 | 1.73 |
| 14 | 1.59 | 2.00 | 1.22 | 1.12 | 1.71 |
| 15 | 2.17 | 1.17 | 5.08 | 2.48 | 1.18 |
| 16 | 3.51 | 2.17 | 1.69 | 1.25 | 4.38 |
| 17 | 1.84 | 0.39 | 3.68 | 2.48 | 0.85 |
| 18 | 1.61 | 2.79 | 4.70 | 2.03 | 1.80 |
| 19 | 1.57 | 1.08 | 2.03 | 1.61 | 2.12 |
| 20 | 1.89 | 2.88 | 2.82 | 2.05 | 3.65 |

The chart designed from the first ten samples is now used for the new data. Again (16) is used to calculate the Bayesian estimates $\hat{x}_{0.99}$ represented by observations 11-20 points in Figure 1a) and the vertical dashed line corresponds to the last in-control sample before the out-of-control state occurs. The chart provides a prompt response at the $7^{th}$ sample after the simulated shift. However, interpreting the OOC scenario expressed in terms of the other involved Weibull parameters, we see that the simulated shift affects the Weibull shape parameter too: specifically, $\beta$ drops from its initial value 4.8 to 2.0. So, this is a typical situation where it would be helpful (and even desirable) to set up a control chart to check for the stability of $\beta$. Figure 1b) shows the obtained chart. As we can see, the change in shape is correctly detected at the $4^{th}$ sample, after the shift. Thus, a warning is raised and investigations for assignable causes can start earlier.

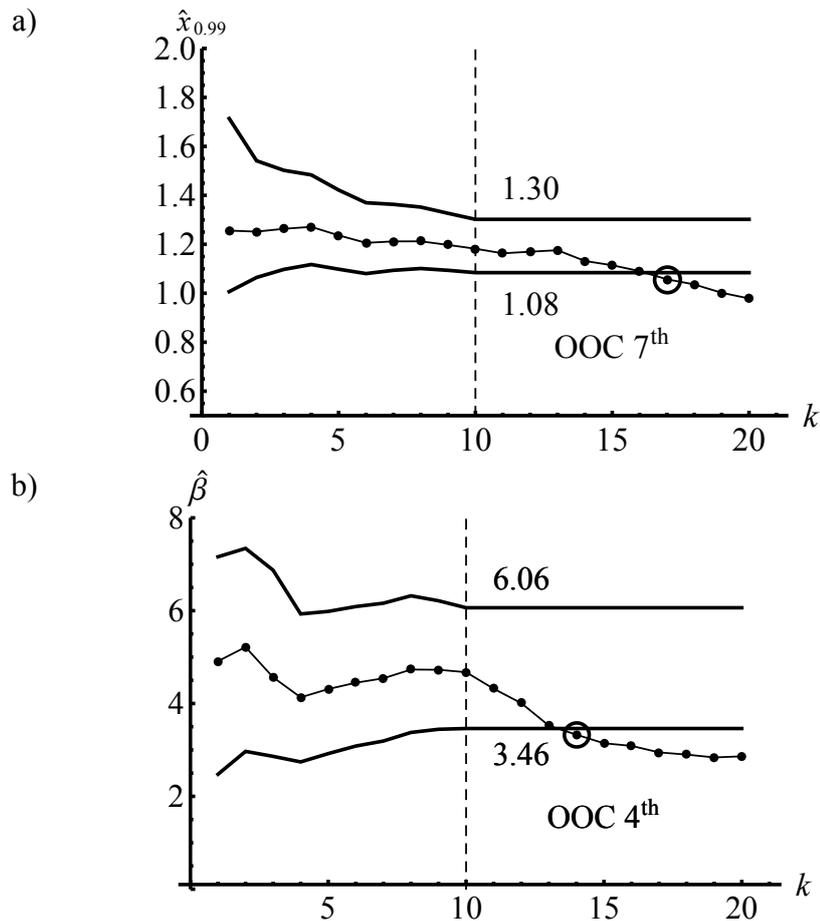

Figure 1. a) Control Chart for $\hat{x}_{0.99}$.    b) Control Chart to Check for $\beta$ Stability.

In order to highlight the considerations done at the end of the previous paragraph, we extended the Phase I to further thirty sample (of size $n = 5$) by resampling from the original set of the in-control data of the first ten rows of Table 1; moreover we included only the last ten samples in the last joint posterior (12). The resulting chart is reported in Figure 2. Having attained a control interval (1.16÷1.27) narrower than before (1.08÷1.30) and having started with a moderately *strong* prior (as before), the chart detects the OOC at $3^{th}$ sample, after the simulated shift, instead that at the $7^{th}$ one.

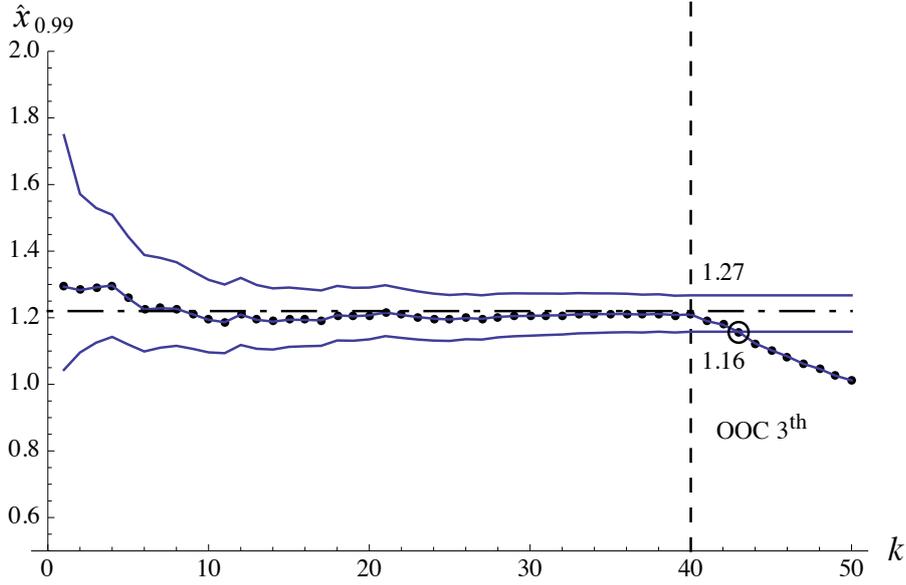

Figure 2. Control Chart of Figure 1a) with a four times longer Phase I.

## 4. Performance of the Control Chart

In this section we present some numerical investigations for exploring the performance of the proposed procedure. Recall that there are two major ideas in the proposed chart: (i) combining prior information with *all* accumulated past and present data, and (ii) continuously *refining* control limits during the Phase I.

First we note that even when poor prior information for $x_R$ and $\beta$ is adopted, the $x_R$ chart performance is not significantly affected. For example, in Figure 3 the prior values are changed to reflect a 25% shift in both directions from the original prior for $x_R$ and $\beta$. This does not merely imply a 25% decrease (increase) in the Weibull process mean $\mu$ and/or variance $\sigma^2$, holding the relationships:

$$\delta = x_R \left[\ln(1/R)\right]^{-1/\beta}; \quad \mu = \delta\ \Gamma(1+1/\beta); \quad \sigma^2 = \delta^2 \left[\Gamma(1+2/\beta) + \Gamma^2(1+1/\beta)\right]. \tag{22}$$

We see that even when the hypothesized prior information is very poor if compared to the baseline (see panel e) in Figure 3), the Bayesian chart still shows a robust diagnostic property, in terms of out-of-control signals (see panels a), c), g) and i) in Figure 3). In fact the number of samples until the control chart signals is always equal to 7. Moreover, the values of the control limits derived from the Phase I are always close to those obtained in the baseline case. Besides, we note that the band width of control limits converges to the same value (i.e., UCL - LCL $\cong 0.22$), regardless of the selected prior values.

We extended our simulated study using a 50% variation too, in both directions from the original prior for $x_R$ and $\beta$, obtaining similar results. In fact, in this case, the number of samples until the control chart signals is always equal to 9 (instead of 7) at most. Moreover, it is worthwhile to note that, when the restriction $\beta_1 + \beta_2 > 2$ is fulfilled, the shift of the $\beta$ prior interval can have moderate effects on the charts performance unlike its even important enlargement. In fact, even using very large $\beta$ prior intervals (e.g.: 1, 10) their effects on the control bands disappears just after that few (say 3) IC data samples are incorporated into $x_R$ and $\beta$ estimates.

The out-of-control (OOC) performance of the proposed control chart was also investigated via a Monte Carlo study to determine if there was an advantage to using the proposed Bayesian approach. The methodology is the same used in Padgett and Spurrier (1990). Specifically, we simulated $N = 1000$ cases of the proposed chart, with different OOC scenarios. For each use of the chart, we performed a Phase I analysis by generating $m$ samples of size $n$ from the in-control (IC) Weibull distribution, using them to compute the control limits.

Then, starting from the subsequent sample $m+1$, we gave the Weibull distribution a percentile shift of a given magnitude, continuing to generate samples until the first point plots outside the control limits. Given an OOC scenario, we calculate the chart performance in terms of the number of samples until the control chart signals, following the process shift. We obtain an estimate of the Average Run Length (ARL) and the Standard Deviation of the Run Lengths (SDRL).

4.1  *First Comparison*

Padgett and Spurrier (1990, see Table 6) report selected OOC ARL values for the Shewhart-type control chart they proposed. Thus, for comparative purposes, we tested the proposed chart in the same scenarios and the results are summarized in Table 2, showing that the chart is able to detect even moderate shifts with small ARLs. As an example, in the third scenario (Table 2, row 3), $R = 0.90$ and the percentile shift is caused by a decrease in the Weibull shape parameter $\beta$ from 3 to 2, leaving the scale parameter $\delta = 1$ unchanged. This change causes an increase of 50% of the standard deviation $\sigma$, but it does not have a significant effect on the process mean $\mu$. Padgett and Spurrier (P&S) (1990) obtain an OOC $ARL = 84.8$. The proposed chart (EP&M) can detect the same shift, with an $ARL = 54.9$ and a Standard Deviation $SDRL = 25.0$. In the last right column of Table 2, we report the OOC ARL for the Weibull control chart presented in Erto and Pallotta (E&P) (2007) that resulted better than other alternative charts on the basis of the study presented in (Hsu Y.C. et al. 2011). As we can see, the new proposed control chart (EP&M) shows even significantly better diagnostic properties.

For this comparison as well as for all the following ones, for $x_R$ we anticipated the corresponding IC value; for $\beta_1$ and $\beta_2$ we anticipated the IC value of $\beta$ multiplied for 0.5 and 1.5 respectively.

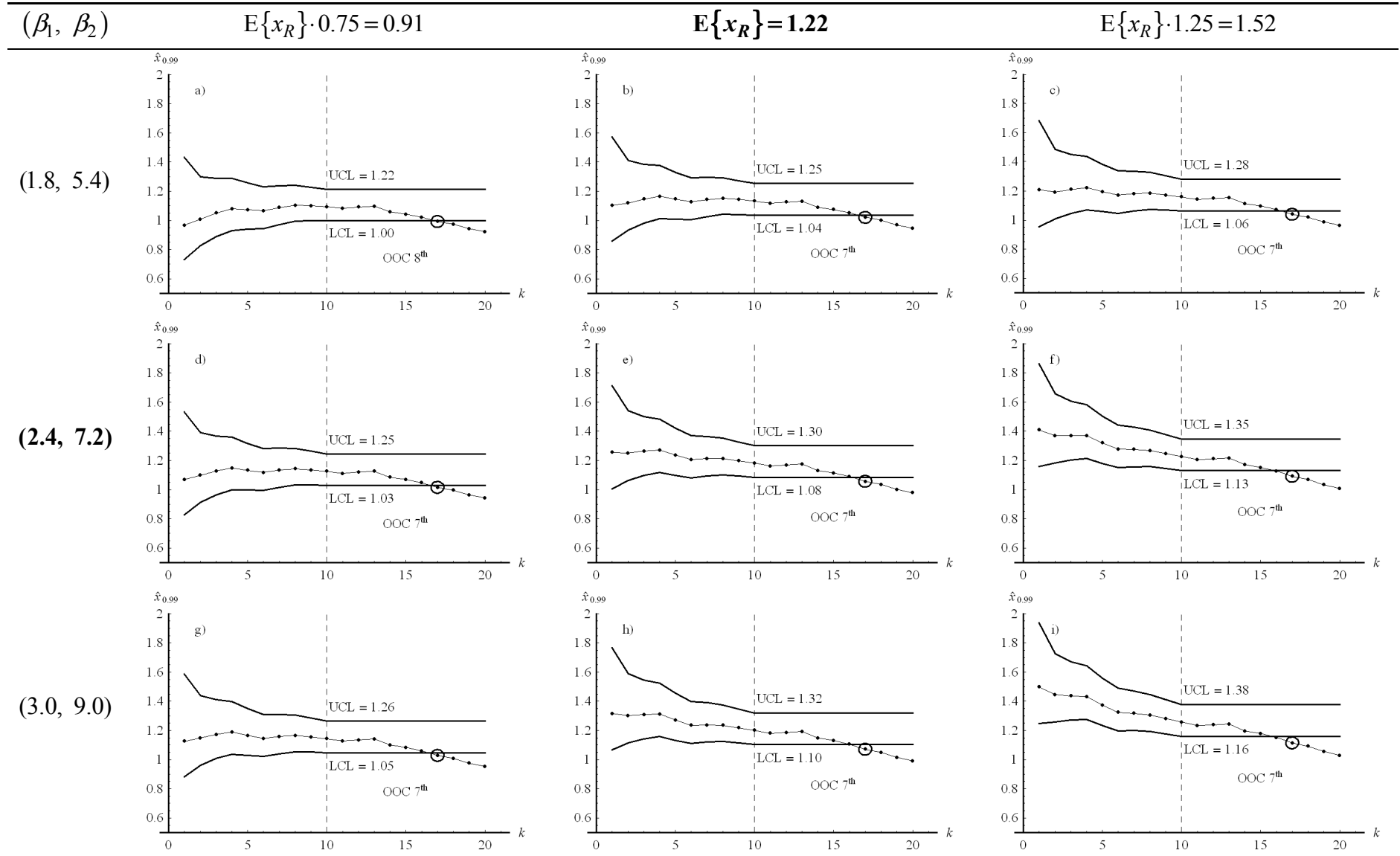

Figure 3. The Effect of Nine Different Priors on the Chart Performance $\left(\hat{x}_{0.99}, n = 5\right)$ (in Boldface the Correct Prior Information).

Table 2. Comparison of the Proposed $x_R$ Chart (EP&M) to Padgett and Spurrier (1990) (P&S) and Erto e Pallotta (2007) (E&P) with $n = 5$ and $m = 25$.

| | IC | | | | | OOC | | | | | | ARL (SDRL) | |
|---|---|---|---|---|---|---|---|---|---|---|---|---|---|
| $R$ | $\mu$ | $\sigma$ | $\beta$ | $\delta$ | $x_R$ | $\mu$ | $\sigma$ | $\beta$ | $\delta$ | $x_R$ | P&S | EP&M | E&P |
| 0.90 | 1.0 | 1.0 | 1.0 | 1.0 | 0.11 | 1.35 | 0.92 | 1.5 | 1.5 | 0.33 | 95.96 | **27.4** (12.8) | 78.8 |
| 0.99 | 0.90 | 0.61 | 1.5 | 1.0 | 0.047 | 1.0 | 1.0 | 1.0 | 1.0 | 0.010 | 42.04 | **26.4** (14.1) | 34.5 |
| 0.90 | 0.89 | 0.32 | 3.0 | 1.0 | 0.47 | 0.89 | 0.46 | 2.0 | 1.0 | 0.32 | 84.82 | **54.9** (25.0) | 66.1 |
| 0.99 | 1.0 | 1.0 | 1.0 | 1.0 | 0.010 | 0.90 | 0.61 | 1.5 | 1.0 | 0.047 | 205.66 | **43.6** (29.4) | 68.4 |

### 4.2 *Stability Check of the Shape Parameter*

Among the monitoring schemes for Weibull processes proposed in literature, Pascual (2010), Pascual and Nguyen (2011), Pascual and Zhang (2011) and Pascual and Li (2012) proposed some control charts to check for the stability of the Weibull shape parameter $\beta$. They argue the strategic role played by this parameter and propose a control chart to check the assumption of stability of $\beta$, which could be tested before looking at the Weibull mean or percentiles. Define $S_\beta = \beta_{OOC}/\beta_{IC}$ the ratio of the OOC $\beta$ value compared to the original IC $\beta$.

Table 3. Out-of-Control Performance of the $x_R$ and $\beta$ Control Charts with IC ARL $= 500$, $n = 1$, $m = 30$, $R = 0.99$.

| | IC | | | | | OOC | | | | | ARL (SDRL) | |
|---|---|---|---|---|---|---|---|---|---|---|---|---|
| $S_\beta$ | $\beta$ | $\delta$ | $x_R$ | $\mu$ | $\sigma$ | $\beta$ | $\delta$ | $x_R$ | $\mu$ | $\sigma$ | Chart $x_R$ | Chart $\beta$ |
| 0.25 | 4.8 | 3.2 | 1.22 | 2.91 | 0.69 | 1.2 | 3.2 | 0.07 | 3.1 | 2.5 | **39.2** (17.2) | 31.4 (18.1) |
| 0.5 | 4.8 | 3.2 | 1.22 | 2.91 | 0.69 | 2.4 | 3.2 | 0.47 | 2.8 | 1.25 | **40.3** (17.2) | 34.2 (19.6) |
| 2 | 4.8 | 3.2 | 1.22 | 2.91 | 0.69 | 9.6 | 3.2 | 1.98 | 3.2 | 0.38 | **60.1** (13.1) | 52.4 (21.3) |
| 4 | 4.8 | 3.2 | 1.22 | 2.91 | 0.69 | 15.2 | 3.2 | 2.5 | 3.1 | 0.22 | **38.1** (10.2) | 36.5 (7.5) |

In Table 3 we report the performances of the proposed control chart for $x_R$ and $\beta$, in terms of OOC ARL and SDRL, with $R = 0.99$. We set an IC ARL $= 500$ and used individual samples. During the Phase I, $m = 30$ training samples were generated. As we can see, the performances of the two proposed charts are comparable, being the chart for $\beta$ slightly more prompt in capturing the instability in $\beta$. Thus, there is essentially no significant difference in terms of the estimated ARLs for the investigated OOC scenarios.

### 4.3 Chart Performance for a Change in the Weibull Percentile

It is important to note that when dealing with Weibull processes, a change in the percentile of interest $x_R$ can be caused by a change in the scale parameter $\delta$ and/or the shape parameter $\beta$. This results in a shift in the Weibull process mean $\mu$ and/or standard deviation $\sigma$, since the relationships (22). As an example, the IC scenario given in the previous section (i.e., $x_{0.99} = 1.22$ and $\beta = 4.8$) corresponds to $\mu = 2.91$ and $\sigma = 0.69$. A percentile shift downward to $x_{0.99} = 0.26$, which is that one simulated in Padgett and Spurrier (1990), can really be achieved via different combinations of the OOC Weibull parameters. Among them, four different OOC scenarios were selected, depending on which Weibull parameters are affected by the change.

Table 4. Out-of-Control Performance of the Control Chart with $m \times n = 50$.

| | Weibull parameter values | | | | | ARL (SDRL) | | |
|---|---|---|---|---|---|---|---|---|
| | $\beta$ | $\delta$ | $x_R$ | $\mu$ | $\sigma$ | $m=10$ $n=5$ | $m=25$ $n=2$ | $m=50$ $n=1$ |
| IC | 4.8 | 3.2 | 1.22 | 2.91 | 0.69 | **370.4** | **370.4** | **370.4** |
| OOC-$\sigma,\beta$ | 1.8 | 3.3 | 0.26 | 2.91 | 1.67 | **6.6** (4.5) | **15.6** (11.7) | **30.5** (23.7) |
| OOC-P&S | 2.0 | 2.6 | 0.26 | 2.26 | 1.18 | **4.5** (1.9) | **13.0** (5.3) | **26.8** (10.7) |
| OOC-$\mu,\beta$ | 2.4 | 1.7 | 0.26 | 1.55 | 0.69 | **2.0** (0.5) | **6.1** (1.3) | **12.9** (2.7) |
| OOC-$\mu,\sigma$ | 4.8 | 0.7 | 0.26 | 0.61 | 0.14 | **1.1** (0.2) | **4.2** (0.5) | **9.1** (1.0) |

In the OOC-$\sigma,\beta$ scenario, the percentile deterioration is caused by an increase in variance and a simultaneous decrease in $\beta$, leaving the mean unchanged. In contrast, the OOC-P&S scenario considered in Padgett and Spurrier (1990), implying both a decrease in mean and an increase in variance. The OOC-$\mu,\beta$ scenario simulates a percentile decrease as a consequence of a decrease in mean and in $\beta$, without changing the variance. Lastly, the OOC-$\mu,\sigma$ scenario describes a decrease in the percentile, leaving the shape parameter $\beta$ unchanged, while both the mean and the variance decrease.

We tested the performance of the proposed charts for different combinations of sample sizes $n$ and numbers of IC samples $m$. Both $n$ and $m$ affect the estimates of the process parameters, and, hence, the chart OOC performance. We first considered a fixed total number of measurements taken to estimate the control limits during Phase I (*i.e.*, $m \times n$ is a constant). This was done in order to highlight the effect of sub-grouping since, as noticed in (Chen, 1997, p. 797), decreasing $n$ and increasing $m$ causes both higher ARL and SDRL.

Table 4 summarizes the results of the chart performance investigation for the four OOC scenarios with IC $ARL = 370.4$. The ARL values are given for samples of size $n = 5$, $n = 2$ and $n = 1$. The corresponding values of the OOC ARL and the SDRL are all based on the same number $m \times n = 50$ of IC data of the Phase I. As we can see, both the OOC ARL and SDRL increase as the sample size decreases in all the investigated OOC scenarios.

Moreover, given any of the OOC scenarios, it is interesting to note that the average amount of data necessary to produce an OOC signal (e.g., $5 \times 6.6 = 33$, $2 \times 15.6 = 31.2$, $1 \times 30.5 = 30.5$) is almost the same, irrespective of the sample size. This is a consequence of the recursive Bayesian estimation feature of the proposed chart.

Table 5. Out-of-Control Performance of the Control Chart with $m = 25$ In-Control samples.

|   |   | Weibull parameter values | | | | | ARL (SDRL) | | |
|---|---|---|---|---|---|---|---|---|---|
|   |   | $\beta$ | $\delta$ | $x_R$ | $\mu$ | $\sigma$ | $m = 25$ $n = 5$ | $m = 25$ $n = 2$ | $m = 25$ $n = 1$ |
|   | IC | 4.8 | 3.2 | 1.22 | 2.91 | 0.69 | **370.4** | **370.4** | **370.4** |
| 1 | OOC-$\sigma, \beta$ | 1.8 | 3.3 | 0.26 | 2.91 | 1.67 | **9.4** (8.3) | **15.6** (11.7) | **23.9** (18.4) |
| 2 | OOC-P&S | 2.0 | 2.6 | 0.26 | 2.26 | 1.18 | **6.9** (2.4) | **13.0** (5.3) | **18.7** (8.3) |
| 3 | OOC-$\mu, \beta$ | 2.4 | 1.7 | 0.26 | 1.55 | 0.69 | **3.5** (0.7) | **6.1** (1.3) | **8.8** (1.9) |
| 4 | OOC-$\mu, \sigma$ | 4.8 | 0.7 | 0.26 | 0.61 | 0.14 | **2.7** (0.5) | **4.2** (0.5) | **6.0** (0.6) |

Then, we investigated the chart performance when the number of the IC samples used in Phase I to estimate the control limits is fixed (i.e., when $m$ is a constant), in order to study the effect of sub-grouping. Table 5 gives the ARL and SDRL values when $m = 25$ for the same OOC scenarios investigated in Table 4. As expected, for fixed $m$, both ARL and SDRL are decreasing functions of $n$. For example, the scenario OOC-$\mu, \beta$ (row 3 in Table 5) corresponds to a decrease in the process mean of about 2 times, while the process standard deviation is kept constant. In this scenario, for $n = 5$, the proposed chart has an OOC $ARL = 3.5$. As another example, the scenario OOC-$\sigma, \beta$ (row 1 in Table 5) corresponds to a more than doubled process standard deviation, while the process mean is kept constant. In this scenario, for $n = 5$ the proposed chart has an OOC $ARL = 9.4$.

In the above OOC-$\mu, \beta$ (row 3 Table 5) and OOC-$\sigma, \beta$ (row 1 Table 5) two scenarios, the classical Shewhart control chart for $n = 5$ would have an OOC $ARL = 6.6$ and an OOC $ARL = 5.0$ in place of 3.5 and 9.4 respectively. However the former (Shewhart) values are merely theoretical, since they hold if the data are normal and the parameters are known. For non-normal data and estimated parameters (as in our case) the actual Shewhart chart ARL values would be greater than these, leading to a performance degradation, as argued in Montgomery (2008, p. 250).

Finally, it must be noted that even when the sample size is $n=1$, the proposed chart has good detection properties in all four scenarios.

## 5. Conclusion

The proposed control chart exploits specific Bayesian estimators for data from Weibull distribution when both parameters are unknown. This is an advantage over other Weibull-based control charts where at least one parameter is assumed to be known. The estimators are called "semi-empirical" since from a technical point of view, the adopted approach can be considered a compromise between pure Bayes and empirical Bayes approach (the former uses prior distributions with *completely specified* hyper-parameters; the latter uses prior distributions estimated using past experimental data, in the context of a two stages sampling plan). In fact, to start the chart it is necessary to anticipate the values of three hyper-parameters (in which prior engineers' knowledge can be really converted) leaving the fourth hyper-parameter *unspecified*. After that, each sampled data set contributes to provide the values of the hyper-parameters of the prior distributions for the analysis of the next sampled data set.

Even in the case of poor priors and small sample sizes, the control chart still has good detection properties and enables prompt decision-making. The chart can be considered to face extremely small samples as in case of short production runs or of low volume production. Moreover, the chart is focused on percentiles, thus giving its operative meaning an explicit role in the design of the control chart. In this way, a more intuitive approach is provided to perform survival monitoring in the Weibull context. Since the proposed chart accumulates information, moderate shifts in the process are readily detected as well. A Monte Carlo study has shown that the chart needs nearly the same *total number* of data to issue an OOC signal irrespective of the sample size. The chart can still work well even starting from a limited number of small samples. In addition, it can be acceptably used for individual observations too. This is an advantage over other published Shewhart-type control charts for Weibull processes available in literature, where the minimum investigated sample size is $n=4$ [Padgett and Spurrier (1990) and Pascual (2010)]. Lastly, the Bayesian framework also provides some additional opportunities such as the computation of the posterior predictive density function that is useful for a process capability analysis.